\documentclass[authorversion,nonacm,sigconf]{acmart}
\usepackage{systeme}
\usepackage{bbm}
\usepackage{tikz}
\usepackage{wrapfig}
\usepackage{caption}
\usetikzlibrary{bayesnet}
\usepackage{dsfont}
\usepackage{comment}
\usepackage{type1cm} %
\usepackage{graphicx} %
\usepackage{xspace} %
\usepackage{balance} %
\usepackage{booktabs} %
\usepackage{multirow} %
\usepackage[font={bf}, tableposition=top]{caption} %
\usepackage[margin={60pt,0pt},position=bottom,subrefformat=simple,labelformat=simple]{subcaption} %
\usepackage{booktabs}

\usepackage{bold-extra} %
\usepackage[vlined,linesnumbered,ruled,noend]{algorithm2e} %
\usepackage{microtype} %
\usepackage{siunitx} %
\usepackage{xfrac} %
\usepackage{mathtools} %
\usepackage{xspace} %
\PassOptionsToPackage{hyphens}{url} %
\PassOptionsToPackage{bookmarks, pdftex, colorlinks=true, pagebackref=true, backref=page}{hyperref} %
\usepackage{cleveref} %
\PassOptionsToPackage{square,numbers}{natbib} %
\usepackage[hyperpageref]{backref} %
\usepackage{hyphenat} %
\usepackage[show]{chato-notes}

\tikzstyle{detobs} = [det, fill = gray!30]

\newcommand{\spara}[1]{\smallskip\noindent\textbf{#1}}
\newcommand{\mpara}[1]{\medskip\noindent\textbf{#1}}

\newenvironment{squishlist}
{\begin{list}{$\bullet$}
 {\setlength{\itemsep}{0pt}
     \setlength{\parsep}{3pt}
     \setlength{\topsep}{3pt}
     \setlength{\partopsep}{0pt}
     \setlength{\leftmargin}{1.5em}
     \setlength{\labelwidth}{1em}
     \setlength{\labelsep}{0.5em} } }
{\end{list}}

\renewcommand*\backref[1]{\ifx#1\relax \else (Cited on #1) \fi}

\newcommand{\bcm}{BCM\xspace}
\newcommand{\bcmfull}{BCM-\texttt{F}\xspace}
\newcommand{\bcmpartial}{BCM-\texttt{P}\xspace}
\newcommand{\bcmnoisy}{BCM-\texttt{N}\xspace}

\newcommand{\likelihood}{\ensuremath{\mathcal{L}}\xspace}
\newcommand{\signs}{\ensuremath{s}\xspace}
\newcommand{\xzero}{\ensuremath{X_0}\xspace}

\newcommand{\relerrormsm}{$89\%$\xspace}
\newcommand{\relerrorpgabm}{$22\%$\xspace}
\newcommand{\timeratio}{200\xspace}

\newcommand{\deltaratioaggregate}{3.1\xspace}
\newcommand{\deltapgabmaggregate}{0.013\xspace}
\newcommand{\deltamsmaggregate}{0.040\xspace}
\newcommand{\timeratiobcmfull}{207.0\xspace}
\newcommand{\timeratiobcmpartial}{11.8\xspace}
\newcommand{\timeratiobcmnoisy}{2.2\xspace}
\newcommand{\timeratioaggregate}{3.9\xspace}
\newcommand{\timepgabmaggregate}{1.7\xspace}
\newcommand{\timemsmaggregate}{7.5\xspace}

\newcommand{\numexperiments}{120\xspace}
\newcommand{\medianXRtwohighT}{0.90\xspace}
\newcommand{\suppmat}{Supplementary Material\xspace}
\DeclareMathOperator*{\argmax}{arg\,max}

\setcopyright{none}
\copyrightyear{2023}
\acmYear{2023}
\acmDOI{XXXXXXX.XXXXXXX}

\acmConference[WSDM '24]{The 17th ACM International Conference on Web Search and Data Mining}{March 04--08,
  2024}{Mérida, Mexico}

\acmBooktitle{Woodstock '18: ACM Symposium on Neural Gaze Detection,
 June 03--05, 2018, Woodstock, NY} 
\acmPrice{15.00}
\acmISBN{978-1-4503-XXXX-X/18/06}

\begin{document}

\title[Likelihood-Based Methods Improve Parameter Estimation in Opinion Dynamics Models]{Likelihood-Based Methods Improve\\Parameter Estimation in Opinion Dynamics Models}

\author{Jacopo Lenti}
\email{jcp.lenti@gmail.com}
\orcid{0000-0003-2886-7338}
\affiliation[obeypunctuation=true]{%
  \institution{CENTAI}, %
  \streetaddress{Corso Inghilterra, 3}
  \city{ Turin}, %
  \country{Italy} %
  \postcode{10138}
}
\affiliation[obeypunctuation=true]{%
  \institution{Sapienza University}, %
  \streetaddress{P.le A. Moro 5}
  \city{ Rome}, %
  \country{Italy} %
  \postcode{00185}
  }

\author{Corrado Monti}
\email{corrado.monti@centai.eu}
\orcid{0000-0001-6846-5718}
\affiliation[obeypunctuation=true]{%
  \institution{CENTAI}, %
  \streetaddress{Corso Inghilterra, 3}
  \city{ Turin}, %
  \country{Italy} %
  \postcode{10138}
}

\author{\texorpdfstring{\nohyphens{Gianmarco~De~Francisci~Morales}}{Gianmarco De Francisci Morales}}
\email{gdfm@acm.org}
\orcid{0000-0002-2415-494X}
\affiliation[obeypunctuation=true]{%
  \institution{CENTAI}, %
  \streetaddress{Corso Inghilterra, 3}
  \city{ Turin}, %
  \country{Italy} %
  \postcode{10138}
}

\renewcommand{\shortauthors}{Lenti, et al.}




\begin{abstract}
We show that a maximum likelihood approach for parameter estimation in agent-based models (ABMs) of opinion dynamics outperforms the typical simulation-based approach.
Simulation-based approaches simulate the model repeatedly in search of a set of parameters that generates data similar enough to the observed one.
In contrast, likelihood-based approaches derive a likelihood function that connects the unknown parameters to the observed data in a statistically principled way.
We compare these two approaches on the well-known bounded-confidence model of opinion dynamics.

We do so on three realistic scenarios of increasing complexity depending on data availability: ($i$) fully observed opinions and interactions, ($ii$) partially observed interactions, ($iii$) observed interactions with noisy proxies of the opinions.
We highlight how identifying observed and latent variables is fundamental for connecting the model to the data. %
To realize the likelihood-based approach, we first cast the model into a probabilistic generative guise that supports a proper data likelihood.
Then, we describe the three scenarios via probabilistic graphical models and show the nuances that go into translating the model.
Finally, we implement the resulting probabilistic models in an automatic differentiation framework (PyTorch).
This step enables easy and efficient maximum likelihood estimation via gradient descent.
Our experimental results show that the maximum likelihood estimates are up to $4\times$ more accurate and require up to $\timeratio\times$ less computational time.
\end{abstract}

\maketitle

\section{Introduction}

\begin{figure}[t]
\centering
\includegraphics[width=\linewidth]{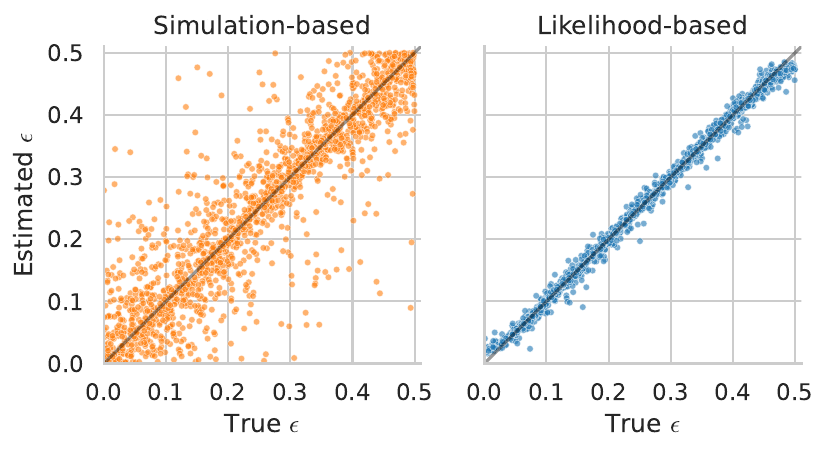}
\caption{Estimation of the `open-mindedness' parameter $\epsilon$ for a Bounded-Confidence Model with a baseline simulation-based approach (Method of Simulated Moments, left) vs a likelihood-based approach (Maximum Likelihood, right). Perfect estimates lie on the diagonal. The likelihood-based approach produces much more accurate estimates.}
\vspace{-\baselineskip}
\label{fig:scatterplot_epsilon}
\end{figure}

The challenges currently facing humanity require global coordination.
The Covid-19 pandemic and climate change are two clear examples of such global problems.
However, people are struggling to find a consensus to tackle these problems, and even simple and effective interventions such as face masks become controversial~\cite{lang2021maskon}. %
Why does this happen?

This question is the main focus of opinion dynamics, the research field that studies the evolution of people's opinions.
And while the literature is replete with possible theoretical models~\cite{lorenz2007continuous,sirbu2017opinion}, few of these have been empirically validated~\cite{platt2020comparison,windrum2007empirical}.
The rise of online social media has provided an unprecedented amount of data about their user's opinions and interactions.
And yet, we are no closer to understanding opinion formation today than 20 years ago.

One of the main obstacles to validating models of opinion formation lies in their nature: agent-based models (ABMs).
A staple of complex systems methodology, ABMs are mechanistic models where \emph{agents} interact among themselves and with an environment according to predefined local rules.
The interactions change the state of the system which in turn affects the interactions in a possibly complex feedback loop, and drive the evolution of the system (e.g., users' opinions).
The system is then simulated to derive outcomes and implications, and to explore what-if scenarios.
The rules are typically \emph{parameterized}, and allow for a variety of agent interactions and system steady states.
The flexible and bottom-up nature of ABMs represents a double-edged sword:
on the one hand, these models are simple to build and understand;
on the other hand, they lack a principled way to use data to estimate their parameters.

Parameter estimation in ABMs, also called \emph{`calibration'} in the literature, is the focus of the current paper.
The vast majority of approaches to the task are \emph{simulation-based}:
they explore the parameter space and run a set of simulations with the parameter values under consideration.
The set of parameters able to produce the data traces ``closest'' to the real data is taken as the result of this procedure.
The definition of closest is arbitrary, and usually consists in comparing summary statistics of the data according to some chosen metric (e.g., differences in mean and variance).
Unfortunately, this procedure can generate quite inaccurate estimates for the parameters of even very simple models.

\Cref{fig:scatterplot_epsilon} shows two scatterplots of the true and estimated parameter values for the well-known bounded-confidence model of opinion dynamics~\citep{deffuant2000mixing,krause2000discrete}.
This model, detailed in \Cref{sec:model}, is the simplest non-linear opinion dynamics ABM, and has a single continuous parameter $\epsilon \in [0,0.5]$ that determines the ``open-mindedness'' of agents. %
Even in this basic example, a simulation-based calibration approach produces relative errors of up to \relerrormsm (90th percentile), as shown in the left plot.
Conversely, the right plot shows the maximum likelihood approach (\relerrorpgabm relative error at 90th percentile).
It is evident that the likelihood-based approach significantly improves on the simulation-based one.
Experiments reported in \Cref{sec:experiments} demonstrate that the likelihood-based method offers more accurate estimates on a variety of different opinion dynamics scenarios.
In addition, this increased accuracy does not come at the cost of increased computational time.
On the contrary, the likelihood-based approach is markedly faster than the simulation-based one. %

In general, the goal of this paper is to compare the likelihood-based parameter estimation to the standard simulation-based solution on a set of scenarios with increasing complexity.
We increase the complexity of the task by taking inspiration from the real challenges that empirical researchers face every day.
We first consider partial observability of the data, e.g., on social media we can only observe `positive' interactions (retweets and likes) by design.
Then, we consider that estimating the exact opinion of users on social media is often impossible, and we have to rely on noisy proxies such as the use of informative hashtags or participation in partisan communities.
These scenarios offer a much more realistic benchmark of the accuracy of parameter estimation for opinion dynamics models.
Our experimental results show that even in the simplest scenario the gap in accuracy between the two methods is surpisingly large.

This paper represents a first step towards a more structured evaluation of the methodological practices in opinion dynamics research.
It emphasizes the importance of aligning our methodologies with the models we use, especially considering the abundant data accessible via social media.
Our aspiration to enhance the understanding of opinion formation hinges on the reliability of our findings.
Given the complexity of the issue, it is evident that building more solid foundations than the current field norms is an essential step forward.

\section{Related Work}
\label{sec:related}
The literature on opinion dynamics is incredibly vast and it would be impossible to review it in this paper, so we limit ourselves to giving a broad overview.
Typically, in opinion dynamics ABMs individuals are represented by agents who share their opinion with their neighbors on an underlying social network, which possibly evolves over time.
The opinion state variable can be discrete~\cite{sznajd2000opinion} or continuous~\cite{deffuant2000mixing}, and represent the stance of a user towards a specific topic, such as politics~\cite{de2022modelling}, vaccines~\cite{alvarez2017epidemic}, or civil rights~\cite{baumann2021emergence}.
The interactions between agents cause the opinion variables to change, thus leading to different scenarios depending on the model in consideration and the specific set of parameters~\cite{peralta2022opinion}.

The advantage of using ABMs for opinion dynamics lies in being able to encode directly individual behaviors and consider only the local properties of the system. 
ABMs are also incredibly flexible, as they can easily incorporate complex strategies or different types of interactions, both with other peers or with the environment.
Thanks to their flexibility, ABMs are used also in several other fields such as economics, ecology, and human mobility~\cite{an2021challenges}.

The interest in opinion dynamics in particular has been recently renewed due to the emergence of phenomena such as polarization, filter bubbles, and echo chambers on social media~\citep{bail2021breaking,garimella2018quantifying,cinelli2021echo,pariser2011filter}.
For instance, the data mining community has produced several interesting models~\citep{chitra2020analyzing,bhalla2023local,abebe2018opinion,musco2018minimizing}.
The availability of data from social media has surely been an incredible catalyst for this type of research.
However, this line of research has not been accompanied by appropriate methodological progress in calibration and validation of the models~\cite{peralta2022opinion}.

\spara{Calibration} is the process of determining the parameters that align a model with an observed dataset~\cite{platt2020comparison}.
Given the generative essence of Agent-Based Models (ABMs), the research community in this realm has commonly embraced simulation-based methodologies as the conventional calibration approach~\cite{cranmer2020frontier}.
These techniques require the execution of numerous simulations of the model, each with a different parameter configuration.
The procedure then selects the parameter configuration that most accurately emulates the data trajectories or system property.
A wide range of approaches have been developed in this direction, which all share three fundamental requisites:
($i$) the identification of summary statistics to quantitatively characterize the model output,
($ii$) the choice of a distance metric to measure the similarity between the summary statistics of the simulated and observed data, and 
($iii$) a parameter sampling procedure aimed at exploring the parameter space~\cite{platt2020comparison, lux2018empirical}.
In addition, these methods require choosing an acceptance threshold to assess whether the obtained simulations have replicated the dataset well enough, and a stopping criterion for the exploration of the parameter space~\cite{hazelbag2020calibration}.

The predominant and de-facto standard simulation-based method is the Method of Simulated Moments (MSM)~\cite{fagiolo2019validation, lux2018empirical}.
In each iteration, this method simulates the ABM to get a time series of generated data.
Then it compares specific moments of the real data and the simulated time series.
After a sufficiently long exploration of the parameter space, the parameters that minimize the discrepancy between the selected moments are chosen as the best estimates~\cite{fagiolo2019validation}.
This approach is also described as reproducing ``stylized facts'', i.e., broad tendencies, in economics.

Researchers have explored different metrics for quantifying the distance between simulated and observed data~\cite{barde2016direct, lamperti2018information, valensise2023drivers}.
The quest for a more efficient parameter space exploration stands as another challenge within simulation-based methods for calibration.
To minimize the number of simulations required, the methods look for the most promising regions of the parameter space.
Several Bayesian strategies integrate a prior knowledge of the distribution of the parameters of interest have been developed.
\citet{grazzini2017bayesian} offer a comprehensive spectrum of approaches, encompassing parametric, non-parametric, and likelihood-free techniques.
Specifically, Approximate Bayesian Computation (ABC) is an established likelihood-free Bayesian technique in common practice.
ABC seeks to maximize the posterior probability of the target quantity, by relying on specific summary statistics, when a likelihood function is not available~\cite{van2015calibration}.
Additionally, several techniques employ deep learning to reduce the number of simulations needed to determine the desired parameters~\cite{dyer2022black, shiono2021estimation}.
In this case, the deep learning model is used as a surrogate for the ABM, so that new data traces can be generated quickly after a more expensive training phase.
Within the stream of Bayesian methods, particle filtering and Sequential Monte Carlo techniques calibrate the parameters online, by simulating the ABM step by step~\cite{lux2023sequential, kattwinkel2017bayesian, ternes2022data, lux2021bayesian}.
Other strategies use simulations to approximate the likelihood~\cite{platt2022bayesian, lamperti2018agent}, or a simulated maximum likelihood estimator~\cite{kukacka2017estimation}.
Particularly notable are the advances in differentiable ABMs, which facilitate the calibration via auto-differentiation techniques and fast tensorized simulations~\cite{quera2023bayesian, chopra2022differentiable}.

However, all the aforementioned approaches share some evident limitations.
First, the arbitrary choices of the summary statistics and of the distance metric heavily affect outcomes of calibration~\cite{fagiolo2019validation}.
Second, the computational cost associated with simulating ABMs presents a substantial hurdle, making iterative runs computationally taxing and challenging to alleviate~\cite{platt2020comparison}.
Consequently, models are often validated solely in simplified versions featuring a limited number of agents~\cite{grazzini2017bayesian}.
Third, these methods suffer from the curse of dimensionality.
As the dimensionality of the parameter space increases, an exhaustive exploration of such space becomes practically unfeasible~\cite{lamperti2018information}.
Moreover, when increasing the complexity of the ABMs, unexpected or unclear behaviors can emerge, which makes purely simulative approaches less reliable~\cite{srikrishnan2021small,lee2015complexities}.
Furthermore, comparing only some summary statistics leads to an inevitable waste of valuable information from the data, thus forcing one to concentrate on a few specific macroscopic properties of the system~\cite{grazzini2017bayesian}.
For these reasons, research in ABMs has recently highlighted the urgency to find new robust methods of calibration for validating existing models, instead of focusing on the definition of new models able to capture different behaviors~\cite{peralta2022opinion, an2021challenges}.

In contrast to simulation-based approaches, only a handful of methodologies employ data to maximize the likelihood of generating them~\cite{dong2016variational, xu2016using, monti2020learning, monti2022learning}.
The goal is always to maximize the likelihood associated with the observed data: \citeauthor{dong2016variational} and \citeauthor{xu2016using} derive the likelihood through stochastic kinetic models, leveraging Variational Inference to optimize it and make predictions.
However, as far as we know, a standardized protocol for implementing these approaches is currently missing, and the solutions presented address specific problems in epidemic models and traffic congestion scenarios.
On the other hand, \citet{monti2020learning,monti2022learning} employ Probabilistic Graphical Models (PGMs) to represent the ABMs.
This novel approach enables the translation of the ABM dynamics into conditional probabilities, encapsulating agent states and interactions as random variables, whether latent or observed.
Within this framework, parameter estimation is not found in the simulations aimed at replicating the desired properties of the data, but the focus shifts to the maximization of the likelihood of generating the observed variables.
Consequently, parameter calibration takes the form of an optimization task, amenable to various automatic differentiation strategies (such as PyTorch~\cite{paszke2019pytorch}, JAX~\cite{jax2018github}, and TensorFlow~\cite{abadi2016tensorflow}).
However, this promising direction has solely delved into specific models, demanding a deeper understanding to reveal its limits and potential.
Beyond the theoretical robustness of maximum likelihood estimation, the questions pertaining to its broader feasibility and empirical outcomes are still open.
In particular, a head-to-head comparison with the simulation-based method serves to explain the real utility of the Maximum Likelihood strategy and its practical viability.

\section{Model and Scenarios}
\label{sec:model}
We consider one of the most well-known opinion dynamics model, i.e., the bounded-confidence model (\bcm) \citep{deffuant2000mixing,krause2000discrete}. %
More specifically, we pick a stochastic version of \bcm inspired by \citet{monti2020learning}.
Our choice is motivated by the popularity of the model, which has been studied extensively, and by its relative simplicity.
Despite being easy to understand and describe this model is the first non-linear model of opinion dynamics ever proposed.
Nevertheless, we shall see that the simulation-based methodology presents significant parameter estimation errors even on such a simple model.

The aim of the model is to represent a situation where the opinions of agents are influenced only by interactions with neighbors that have similar opinions.
More specifically, an agent is influenced by an interaction with another agent only if their opinions are within their \emph{confidence interval}, encoded in a parameter $\epsilon$.
These interactions drive the evolution of their opinions by mutual influence and convergence.
This model is an ABM: each agent behaves according to independent, local rules.

Formally, the model operates as follows.
We consider a population $V$ of $N$ agents.
At time $t$, each agent $u$ has an opinion $x^u_t \in [0,1]$, with initial opinion $x_0^u$ drawn uniformly at random.
At each time $t \in \{1, \ldots, T\}$, $m$ pairs of agents are extracted uniformly at random from $V \times V$.
We indicate these extractions with $e_t$.
An interaction occurs for each pair of agents in $e_t$, for a total of $M = m \cdot T$ interactions.

In the original formulation of \bcm, when $u$ and $v$ interact, if the distance between their opinions is lower than $\epsilon$ (the bounded-confidence parameter), their opinions converge towards their average at a rate $\mu$ (convergence parameter).
Otherwise, their opinions remain unchanged.
In the stochastic version we consider, each outcome happens with some probability.
For each interaction $(u, v)$ at time $t$, a sigmoid function defines the probability of that interaction being positive.
Formally, we extract a sign $s_{t}^{u, v} \in \{0, 1\}$ as a Bernoulli trial
\begin{equation}
    P(s_{t}^{u, v} = 1) = \sigma \Big( 
        \rho \cdot (\epsilon - \lvert x^u_t - x^v_t \rvert) 
    \Big)\,,
    \label{eq: kappa}
\end{equation} where $\sigma(x) = \frac{1}{1 + e^{-x}}$ is the logistic function and $\rho$ is a steepness parameter that tunes the amount of stochastic behavior.
Then, the agents update their opinions as follows:

\begin{equation}
  \left\{
    \begin{aligned}
    x^u_{t+1} &= x^u_t + s_{t}^{u, v} \cdot \mu (x^v_t - x^u_t) \\
    x^v_{t+1} &= x^v_t + s_{t}^{u, v} \cdot \mu (x^u_t - x^v_t)\,,
    \end{aligned}
  \right.
  \label{eq:BC_update}
\end{equation} that is, after an interaction agents are likely to get closer if they were within their \emph{confidence boundary}, i.e., if $\lvert x^u_t - x^v_t \rvert < \epsilon$. 
Note that for $\rho \to \infty$ the model is equivalent to the original, deterministic \bcm model~\citep{monti2020learning}.

\mpara{Scenarios.}
The goal of this work is to compare two different paradigms for the estimation of parameters: likelihood-based methods based on probabilistic generative models, and simulation-based methods, i.e., simulated moments.
To compare these two paradigms, we focus on the estimation of $\epsilon$, since it is the most characteristic parameter of the model.
Recapping, the variables involved are thus the following:
\begin{squishlist}
    \item $\epsilon$ is the bounded-confidence parameter of the \bcm model, it is the target of the estimation.
    \item $\mu$ is the convergence parameter, it is known.
    \item $X_t = (x^1_t, \ldots, x^N_t) \in [0,1]^N$ the opinions of the agents at time $t$. It depends on the opinions and interactions at previous time steps. The initial state \xzero is extracted from a uniform distribution in $[0,1]^N$.
    \item $e_t = (e^1_t, \ldots, e^m_t) \in (V^2)^m$ are the pairs of agents extracted at time $t$. They are extracted uniformly at random (with replacement) from $V^2$. A single pair of agents is an interaction, and it is defined as $e^i_t = (u,v)$.
    \item $\signs_t \in \{0,1\}^m$ are the sign of interactions at time $t$, that is, whether each interaction has resulted in opinion change. %
    Sign $s^i_t$ is the outcome of the interaction $e^i_t$. If $s^i_t = 1$ the interacting agents update their opinions according to the \bcm rules, otherwise $s^i_t = 0$. 
    The outcomes $\signs_t$ of the interactions depend on the agents involved $e_t$, their opinions $X_t$, and the parameter $\epsilon$.
\end{squishlist}

\smallskip

To evaluate the performances of the two estimation methods under a variety of conditions, we run experiments in different scenarios with varying knowledge about the opinion formation process.
Specifically, we compare three different scenarios:
    (i)~fully-observed interactions~(\bcmfull);
    (ii)~partially-observed interactions~(\bcmpartial);
    (iii)~noisy proxies for opinions~(\bcmnoisy).
Let us now give the details of these three scenarios.

\spara{Fully-observed Bounded Confidence (\bcmfull).}
In the first scenario, we assume complete knowledge of the other variables in the process, which means that we observe the initial condition \xzero, the identities of the pairs of interacting agents $e_t$, and the resulting outcome of the interaction $\signs_t$.
Given the observations of all of these variables, the task is always to estimate the unknown parameter~$\epsilon$.
The results in \Cref{fig:scatterplot_epsilon} are obtained from this scenario.

\spara{Partially-observed Bounded Confidence (\bcmpartial).}
In our second scenario, we reduce the available knowledge: we can only observe interactions with a positive outcome.
This scenario is realistic in many cases where data is available: for instance, retweets on Twitter represent support, but obtaining data about disagreeing users is much harder.
Hence, we assume to observe the opinions of the agents $X_t$, and the pairs of interacting agents having a positive interaction: $e_t$ is observable only when $\signs_t = 1$.

\spara{Bounded Confidence with Noisy Opinions (\bcmnoisy).}
The third scenario assumes having only partial and noisy information on the opinions while observing all the interactions $e$ and their signs \signs.
We introduce an observed variable $Y_t  \in \{0,1\}^k$ which is a proxy of the opinions of a subset $k$ of the agents.
To obtain this proxy, at each time $t$ a set of $k$ agents is selected uniformly at random to share some content with label $y^u_t \sim \text{Bernoulli}(x^u_t)$.
In total, we have $K = k \cdot T$ binary proxies.
The opinion vector \xzero is instead latent, as well as its evolution $X_t$.
These conditions fit computational social science studies, where the exact stance of social media users is hard to know, but we have some proxies of their opinions in the form of the hashtags they use or the communities they participate in.
Similarly, binary labels of the stance of a user are easier to elicit from human labelers than their continuous hidden representation.

\section{Likelihood-based Estimation} %
\begin{figure*}[t]
\begin{center}
    \begin{subfigure}{0.25\textwidth}
      \begin{tikzpicture}
      \scalebox{0.65}{
        \node[det]                          (X)     {$\mathbf{X}_t$};
        \node[obs, above=0.4cm of X]     (X0)    {$\mathbf{X}_0$};
        \node[det, below=2.8cm of X]        (Xt1)   {$\mathbf{X}_{t+1}$};
        \node[obs, below=1.8cm of X, xshift = - 1.8cm]        (s)     {$s$};
        \node[const, below=0cm of X, xshift=-1.8cm]     (e)     {$e$};
        \node[latent, below=0.2cm of e, xshift=-1.6cm]     (eps)     {$\epsilon$};

        \plate [minimum size=3cm, minimum width=3.8cm] {} {(s)
        (e) (X) (Xt1) } {$T$} ;

        \edge {e} {s}
        \edge {X} {s}
        \edge {X0} {X}
        \edge {X} {Xt1}
        \edge {s} {Xt1}
        \edge {eps} {s}
      }
      \end{tikzpicture}
      \vspace{-20mm}
      \caption{\bcmfull}
      \label{fig:BC_full}
    \end{subfigure} %
    \begin{subfigure}{0.25\textwidth}
      \begin{tikzpicture}
        \scalebox{0.65}{
          \node[det]                          (X)     {$\mathbf{X}_t$};
          \node[obs, above=0.4cm of X]     (X0)    {$\mathbf{X}_0$};
          \node[det, below=2.8cm of X]        (Xt1)   {$\mathbf{X}_{t+1}$};
          \node[obs, below=1.8cm of X, xshift = - 1.2cm]        (s)     {$s$};
          \node[latent, below=0cm of X, xshift=-1.8cm]     (e)     {$e$};
          \node[detobs, below=0.4cm of s, xshift=-1cm]     (ehat)     {$\hat{e}$};

          \node[latent, below=0.2cm of e, xshift=-2cm]     (eps)     {$\epsilon$};
          \plate [minimum size=3cm, minimum width=4.1cm] {} {(s)
          (e) (X) (Xt1) (ehat) } {$T$} ;

          \edge {e} {s}
          \edge {X} {s}
          \edge {X0} {X}
          \edge {X} {Xt1}
          \edge {s} {Xt1}
          \edge {eps} {s}
          \edge {s} {ehat}
          \edge {e} {ehat}
          \edge {ehat} {Xt1}
        }
      \end{tikzpicture}
      \vspace{-20mm}
      \caption{\bcmpartial}
      \label{fig:BC_partial}
    \end{subfigure} %
    \begin{subfigure}{0.25\textwidth}
      \begin{tikzpicture}
        \scalebox{0.65}{
          \node[det]                          (X)     {$\mathbf{X}_t$};
          \node[latent, above=0.4cm of X]     (X0)    {$\mathbf{X}_0$};
          \node[det, below=2.8cm of X]        (Xt1)   {$\mathbf{X}_{t+1}$};
          \node[obs, below=1.3cm of e]        (s)     {$s$};
          \node[const, below=0cm of X, xshift=-1.8cm]     (e)     {$e$};
          \node[obs, below=0.6cm of X, xshift=1cm]     (Y)     {$Y_t$};
          \node[latent, below=0.2cm of e, xshift=-1.4cm]     (eps)     {$\epsilon$};
          \plate [minimum size=3cm, minimum width=4.1cm] {} {(s)
          (e) (X) (Xt1) (Y) } {$T$} ;

          \edge {e} {s}
          \edge {X} {s}
          \edge {X0} {X}
          \edge {X} {Xt1}
          \edge {s} {Xt1}
          \edge {eps} {s}
          \edge {X} {Y}
        }
      \end{tikzpicture}
      \vspace{-20mm}
      \caption{\bcmnoisy}
      \label{fig:BC_noisy}
    \end{subfigure} %
    \hspace{60pt}

\caption{Representation of the Bounded Confidence Model as a probabilistic graphical model in three different scenarios. In the first scenario \subref{fig:BC_full} \bcmfull, both the initial state \xzero and the signs \signs are fully-observed. In \bcmpartial \subref{fig:BC_partial}, \xzero is still known but only the positive interactions are observed, represented by the fact that only some of the edges $\hat{e}$ are observed (chosen deterministically). In the final one \bcmnoisy \subref{fig:BC_noisy}, the initial state \xzero is unknown, and at each timestamp we get a random noisy proxy $Y_t$ of the state of some of the agents.}
\label{fig:graphical_models}
\vspace{-\baselineskip}
\end{center}
\end{figure*}
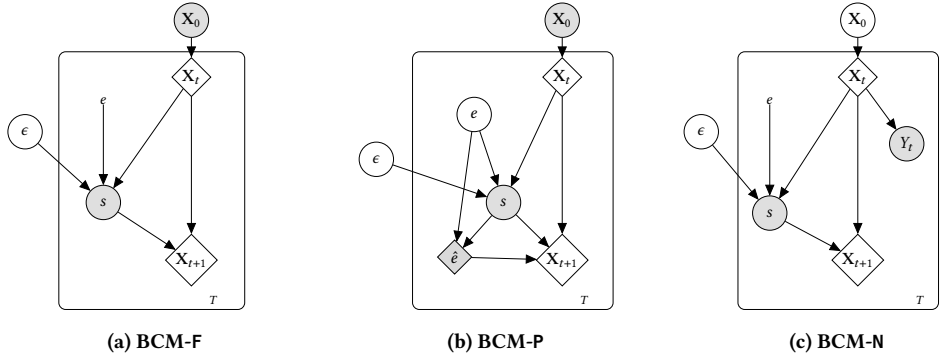

The goal of this work is to compare a likelihood-based method, and in particular one based on maximum likelihood (ML) estimation via gradient-based optimization, against a simulation-based one (method of simulated moments, MSM).
As explained in~\Cref{sec:related}, MSM simulates the ABM while exploring the space of the unknown parameters.
The parameter that generated the simulated trajectory closest to the observed one is the final parameter estimate $\hat{\epsilon}$.

Conversely, for ML, we maximize the likelihood function for the parameter estimate $\hat{\epsilon}$ after having cast each ABM into an equivalent probabilistic generative agent-based model \cite{monti2020learning, monti2022learning}.
To maximize the likelihood function we need to formalize the conditional probabilities connecting the parameters, the latent variables, and the observed data.
Often, the locality of agent-based models allows to efficiently capture conditional (in-)dependencies between random variables, thus making it easier to treat the likelihood.
Then, we estimate the target parameter by maximizing the marginal likelihood of the parameters of interest from the observed data $z$, i.e., $\hat{\epsilon} = \argmax_\epsilon \likelihood(\epsilon \mid z)$.

In the remainder of this section, we analyze each of the scenarios considered, show how it gets represented as a probabilistic graphical model (PGM), and derive its marginal likelihood w.r.t. the parameter of interest $\epsilon$.
\Cref{fig:graphical_models} depicts the PGMs of the three scenarios.
Circles represent stochastic variables, diamonds deterministic variables, and the known parameters have no shape.
($\mu$ and $\rho$ are omitted for simplicity).
Grey colored shapes are observed variables, and white ones are latent.
$X_t$ is known in \bcmfull and \bcmpartial, because it is a deterministic function of observed variables.

\subsection{Fully-observed Scenario (\bcmfull)}
\label{sec:simple_BC}

The PGM induced by the model in this scenario is represented in \Cref{fig:BC_full}.
Our goal is to optimize the marginal likelihood of $\epsilon$ given the observed data.
Since the only observed variable that depends on the latent parameter is the sequence of interaction outcomes \signs, we write the likelihood function as

\begin{equation}
        \likelihood(\epsilon) = P(s \mid \epsilon).
    \label{eq: likelihood_full1}
\end{equation}

Note that $X_t$ is known for any $t$, as it is a deterministic function of the interaction signs \signs and of \xzero, which are both observed in this scenario.
Thanks to the conditional independencies found in the PGM, \Cref{eq: likelihood_full1} factorizes over the sequence of interactions as

\begin{equation*}
    P(s \mid \epsilon)
    = \prod_{j = 1,\ldots,M} P(s^j \mid X_t, \epsilon).
\end{equation*}

Finally, we simply derive the likelihood of a Bernoulli distribution from \Cref{eq: kappa},
\begin{align}
    P(s^j \mid X_t, \epsilon) = s^j \kappa_{\epsilon} (e^j) + (1 - s^j) (1 - \kappa_\epsilon(e^j)),
\end{align}
where for brevity we indicate with $\kappa_{\epsilon} (e^j)$ the sigmoid as a function of the pair of agents $e^j = (u, v)$, that is, \begin{equation*}
    \kappa_\epsilon(e^j) = \sigma \Big( \rho \cdot (\epsilon - \lvert x^u_t - x^v_t \rvert) \Big).
\end{equation*}

By switching to the log-likelihood for optimization purposes (with abuse of notation), we obtain the following objective function for the PGABM
\begin{equation}
    \likelihood(\epsilon)
    = \sum\limits_{j = 1,\ldots,M} \log \left( s^j \kappa_{\epsilon} (e^j) + (1 - s^j) (1 - \kappa_\epsilon(e^j)) \right).
    \label{eq:simple_BC}
\end{equation}

We get an estimate $\hat{\epsilon}$ simply by maximizing this function with gradient descent.

\subsection{Partially-observed Scenario (\bcmpartial)}
\label{sec:observed_positive}

In this scenario, we only have knowledge of which agents interacted if their interaction caused opinion change.
Therefore, the pair of interacting agents $e$ is, in the general case, latent.
We introduce $\hat{e}^j$ as an observed deterministic variable, that is equal to $e^j$ if $s^j = 1$, and is equal to $\varnothing$ if $s^j = 0$
(where the symbol $\varnothing$ denotes a null observation).
The induced PGM is represented in \Cref{fig:BC_partial}.
Again, our goal is to estimate the bounded-confidence parameter $\epsilon$ from the observable data.
That is, we want to maximize the marginal likelihood w.r.t. the parameter $\epsilon$ given \signs and $\hat{e}$. 
Similarly to the previous scenario, we have knowledge of \xzero and of the interactions that cause opinion change (which are only the positive ones according to the BCM), and therefore we fully observe $X_t$.
When considering the likelihood, we can again factorize each element of the sequence of interactions
\begin{equation*}
    \likelihood(\epsilon) = \log P(\hat{e}, s\mid \epsilon) =
    \sum_{j = 1,\ldots,M} \log P(\hat{e}^j, s^j \mid \epsilon)
\end{equation*}

In this scenario, we do not know which agents interacted when the interaction does not change opinions.
Thus, we have to integrate over all the possible extractions of $e$, which is drawn uniformly from $V^2$.
The probability above further factorizes as
\begin{align*}
 P(\hat{e}^j, s^j \mid \epsilon)
& = \sum\limits_{e^j \in V^2} P(\hat{e}^j, s^j, e^j \mid \epsilon)\\
& = \sum\limits_{e^j \in V^2} P(\hat{e}^j \mid s^j, e^j, \epsilon)
                        P(s^j  \mid e^j, \epsilon) P(e^j \mid \epsilon).
\end{align*}
Let us now look at these three probabilities.
Since $\hat{e}^j$ is deterministic given $s^j, e^j$, the first one is equal to $1$ when $s^j = 0$ $\land$ $\hat{e}^j = \varnothing$, or when $s^j = 1$ $\land$ $\hat{e}^j = e^j$, and is equal to 0 otherwise.
The second one, $P(s^j  \mid e^j, \epsilon)$ is given by the sigmoid, and we write it as $\kappa_{\epsilon} (e^j)$ when $s^j = 1$ and $(1 - \kappa_\epsilon(e^j))$ otherwise.
Finally, $P(e^j \mid \epsilon)$ is constant with respect to $\epsilon$ and therefore we can ignore it for our estimation task.
Therefore, we can divide interactions according to their sign $s^j$ and rewrite the log-likelihood as
\begin{equation*}
\likelihood(\epsilon) =
    \sum\limits_{j: s^j = 1} \log \left( \kappa_{\epsilon}(\hat{e}^j) P(e^j) \right)  +  \sum\limits_{j: s^j = 0} \log \left( \sum\limits_{e^j \in V^2} (1 - \kappa_{\epsilon}(e^j)) P(e^j) \right),
\end{equation*}
which is our objective function in this scenario.
In practice we do not need to sample all the possible pairs of agents, as we can approximate $\sum\limits_{e^j \in V^2} (1 - \kappa_{\epsilon}(e^j)) P(e^j)$ with a sample mean. 

\subsection{Noisy Opinions Scenario (\bcmnoisy)}
\label{sec:bcmnoisy}

\Cref{fig:BC_noisy} shows the corresponding PGM.
Differently from the previous scenarios, here \xzero is latent, and therefore we do not know $X_t$. %
Even though we wish to estimate only $\epsilon$, we need to also estimate \xzero (and thus $X_t$) as the two variables are not independent given the signs \signs.
Since gradient descent relies on backpropagation to update the estimate of \xzero, we write the function $X_t = f_t(\xzero)$ as the following efficient matrix operation.
Let $A_t$ be the adjacency matrix of positive interactions at time $t$, that is $A_{i,j} = 1$ iff $\exists j : e^j = (i,j) \wedge s^j = 1$.
Then, we can write \Cref{eq:BC_update} as
\begin{equation*}
    X_{t+1} = X_t + \mu \left( \left(X_t \textbf{1}^T - \textbf{1} X_t^T\right) \odot A_t\right) \textbf{1}^T ,
    \label{eq: tensorized_update}
\end{equation*}
where $\odot$ indicates the Hadamard product.

The log-likelihood function of this model can be divided into two factors, one for the interaction signs \signs and one for the opinion proxies $Y$, since they are independent given $\xzero, \epsilon$
\begin{align*}
    \likelihood(\epsilon) &= \log P(s, Y \mid \epsilon, \xzero) = \log P(s \mid \epsilon, \xzero) + \log P(Y \mid \epsilon, \xzero) \\
    &= \sum\limits_{j = 1,\ldots,M} \log \left( s^j \kappa_{\epsilon} (e^j) + (1 - s^j) (1 - \kappa_\epsilon(e_t)) \right) \nonumber\\
    &+ \sum\limits_{j = 1,\ldots,K} \log \left( y_j f_t(\xzero)_j + (1 - y_j) (1 - f_t(\xzero)_j) \right) .
    \label{eq: loss_evidences}
\end{align*}
In the optimization procedure, we initialize \xzero randomly and use the update rule $f$ to maximize this objective function.

\section{Experiments}
\label{sec:experiments}

We compare the accuracy in parameter estimation of the simulation- and likelihood-based methods with a set of experiments with equal data availability.
For all our experiments, the ground-truth model used to generate the data traces and the generative model assumed for both methods are the same, i.e., there is no misspecification error.
For each experimental configuration, we generate \numexperiments data traces by varying the seed of the pseudorandom generator.
In all the experiments we fix $N = 100$, $\rho = 16$, $\mu = 0.1$.
We vary $T \in [16, 32, 64, 128, 256, 512]$ and $m \in [1, 4,8,16]$.
In \bcmnoisy we also vary $k \in[4,8,16]$.
These parameters are available to both methods and the only unknown one is $\epsilon$.
The code to reproduce these results is open-source and available at \url{https://anonymous.4open.science/status/ABMcalibration-MLvsMSM-4837}.

\begin{figure*} 
\centering
\includegraphics[width=1.\linewidth]{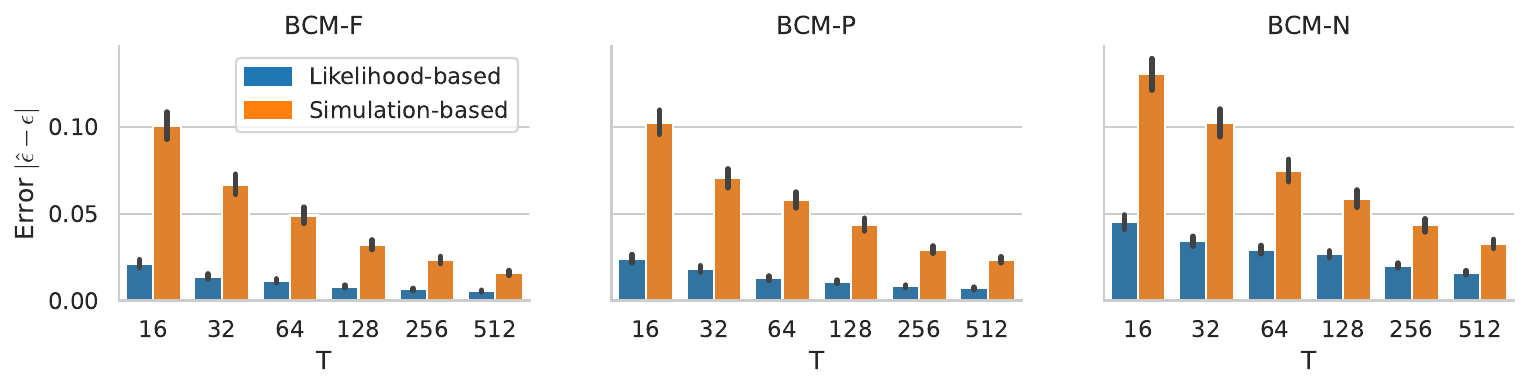}
\caption{
Average absolute error of estimates of $\epsilon$ with Maximum Likelihood and Method of Simulated Moments as a function of $T$.
Error bars represent the 95\% confidence intervals.
The likelihood-based approach produces estimates with several times smaller errors.
}
\label{fig:abs_error_epsilon}
\end{figure*}

\begin{figure*} 
\centering
\includegraphics[width=1.\linewidth]{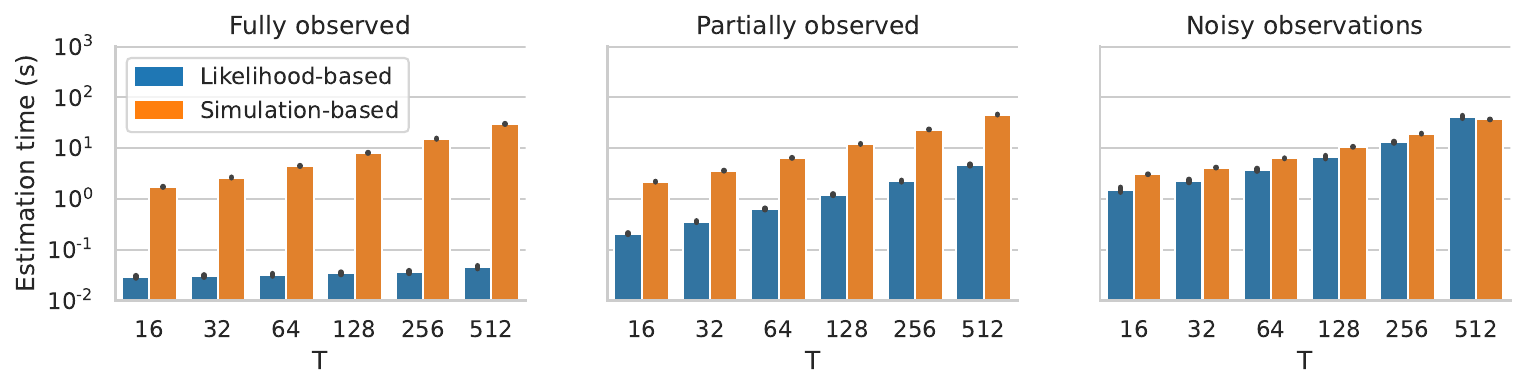}
\caption{
Average time in seconds required to estimate $\epsilon$ with Maximum Likelihood and Method of Simulated Moments as a function of $T$.
Error bars represent the 95\% confidence intervals.
The likelihood-based approach improves the computational cost by orders of magnitude when it does not need to reproduce the entire opinion trajectory (left and center plots).
In the last scenario (right), at each iteration the entire opinion trajectory $X$ needs to be re-computed from the new estimate of \xzero.
}
\label{fig:estimation_time}
\end{figure*}

\spara{MSM parameter settings.}
For the implementation of the MSM we rely on Black-it~\cite{benedetti2022blackit}, a recent, open-source, simulation-based calibration framework.
As pointed out in \Cref{sec:related}, MSM requires some arbitrary choices:
\begin{squishlist}
    \item \emph{Summary statistics}. We compute the time series $w$ of the number of interactions with positive outcomes ($w_t = \sum_j \signs_t^j$).
    Following previous literature on MSM~\cite{benedetti2022blackit}, we define a vector of 18 moments of $w$ (first 4 moments and the first 5 auto-correlations of $w$, first 4 moments and first 5 auto-correlations of the absolute differences at lag one of $w$).
    \item \emph{Loss function}. It incorporates the notion of distance between the simulated and observed time series of the ABM. We use the Mean Squared Error (MSE) between the simulated and observed summary statistics.
    \item \emph{Exploration strategy}. We use the Halton sampler~\cite{kocis1997computational} to explore the parameter space, as it achieves the best results in tuning.
    \item \emph{Number of simulations}. This one is the main tuning hyperparameter for MSM: since the estimation time increase linearly with the number of simulations, there is a time/accuracy trade-off.
    When tuning this parameter, we observe an increase in accuracy until $200$ simulations, and a knee corresponding to such value.
    Hence, in all the experiments we run $200$ simulations (details on the tuning are shown in \suppmat).
\end{squishlist}

In addition, some small customization to each scenario is required to run MSM.
\begin{squishlist}
    \item \bcmfull: at each timestep, the signs $\signs_t$ are simulated from the known agents $e_t$ and opinions $X_t$.
    \item \bcmpartial: in this case, the agents $e_t$ are latent. However, simulating $\signs_t$ from the observed $\hat{e}_t$ would end up in an upward-biased estimate of $\hat{\epsilon}$, since the sample $\hat{e}_t$ represent pairs of agents that have had a positive interaction, and are thus likely to be close each other in opinion space. Hence, at each step, we sample $e_t \in V^2$ and simulate $\signs_t$ by knowing $X_t$.
    \item \bcmnoisy: since \xzero is latent, a sample from a uniform distribution is used as its estimate $\hat{\xzero}$. At each timestep, the signs $\signs_t$ are simulated from $e_t$ and from the current estimate of $\hat{X}_t$ (as a function of $\hat{X}_0$, $e_{\tau < t}$, $s_{\tau < t}$).
    Note that MSM does not provide a way to estimate latent variables $X_t$, but only parameters such as $\epsilon$.
    Moreover, to consider the information from $Y_t$, we sample a set of $k$ agents $\Tilde{V}_t$, and sample $\hat{Y}_t \sim \text{Bernoulli}(X_t(\Tilde{V}_t))$.
    In this scenario, we compute the moments of three different time series: $w_t = \sum_j \signs^j_t$, $\gamma^0_t = \mathbbm{E} (\hat{Y}_t)$ and $\gamma^1_t = \mathbbm{V} (\hat{Y}_t)$.
    The distance between the observed and simulated time series is the sum of the MSEs between the 18 moments of $w$, $\gamma^0$, and $\gamma^1$.
\end{squishlist}

\spara{ML parameter settings.}
We maximize the objective functions resulting from each scenario by using gradient descent on PyTorch, a Python package that enables automatic differentiation.
Following hyperparameter tuning, we choose distinct optimizers for the three scenarios.
Notably, and differently from MSM, in \bcmnoisy the optimization process involves the estimation both of the target parameter $\epsilon$ and the latent variable \xzero.

The specific optimizers and their respective learning rates for each model are as follows:
\begin{squishlist}
\item    \bcmfull: Nadam with a learning rate of \num{0.1}.
\item    \bcmpartial: Adam with a learning rate of \num{0.05}.
\item    \bcmnoisy: RMSprop with learning rates for $\epsilon$ set to \num{0.05} and for \xzero set to \num{0.5}.
\end{squishlist}
All optimizers are stopped after \num{200} epochs if they have not converged yet.
For further details on the tuning of these parameters refer to the \suppmat.

\spara{Implementation details.}
Some implementation tricks are used in the optimization pipeline of our likelihood functions:
\begin{squishlist}
    \item To constrain the parameter space ($[0,0.5]$ for $\epsilon$ and $[0,1]$ \xzero), we represent the target parameters as the logit function of an unbounded optimization variable $\theta \in \mathbb{R}$. For example, $\hat{\epsilon} =  \sigma(\theta) / 2$.
    This choice avoids the need for clipping and possible problems with vanishing gradient.
    \item To estimate the parameters of \bcmpartial we need to compute the population mean, which involves averaging $\kappa_{\epsilon}(e)$ at each step for all possible agent pairs $e$.
    To reduce the computational cost of this process, we employ a sample mean of $\kappa_{\epsilon}(e)$ instead.
    \item In \bcmnoisy \xzero is latent, and the likelihood function is symmetric around the midpoint of the opinion space (\num{0.5}), i.e., \xzero and $1-\xzero$ are both equally good estimates for the latent opinions.
    This behavior arises from the arbitrariness of the two endpoints of the opinion scale, in addition to the fact that the likelihood computed from the interactions dependent only on the distance between agent opinions, not their absolute values. 
    We implement an approach to circumvent encountering a local optimum when $\hat{X}_0 \approx 1 - \xzero$.
    During each epoch, we calculate the objective function as $\Tilde{\likelihood}(\epsilon, \xzero \mid \signs, Y) = \max(\likelihood(\epsilon, \xzero \mid \signs, Y), \likelihood(\epsilon, 1 - \xzero \mid \signs, Y))$.
    Consequently, we select the value of $\hat{\xzero}$ that optimizes this modified objective function.
\end{squishlist}

\mpara{Results.}
\Cref{fig:abs_error_epsilon} shows the average absolute errors $\delta_{ML} = \lvert \hat{\epsilon}_{ML} - \epsilon \rvert$ and $\delta_{MSM} = \lvert \hat{\epsilon}_{MSM} - \epsilon \rvert$ in the three scenarios, as a function of $T$ (and relative $95\%$ confidence intervals).
The difference in the performances of the two methods is evident.
Aggregating over the three scenarios, we obtain a median decrease in the absolute error of $\deltaratioaggregate\times$ (median $\delta_{ML}$  \deltapgabmaggregate vs $\delta_{MSM}$ \deltamsmaggregate).
Considering that $\epsilon$ has maximum value \num{0.5}, an average error of \num{0.1} as obtained by MSM in low-data configurations represents a minimum relative error of $20\%$.
The error of MSM increases slightly on more complex scenarios, although not in a significant way.
Instead, the ML approach achieves very low errors for the fully-observed scenario, and only marginally higher errors for the more complex scenarios.
Reassuringly, both methods are able to improve their estimates when using more data.

\Cref{fig:estimation_time} compares the computational costs of the estimation (note the logarithmic axis).
In \bcmfull and \bcmpartial the comparison is strikingly one-sided.
MSM requires repeating the whole simulation of the ABM hundreds of times, thus incurring a significant computational burden.
However, in \bcmnoisy the Maximum Likelihood method shows similar estimation time.
This increase in computational cost is due to the need to re-compute $X$ from the current estimate of \xzero at each epoch.
The median ratios between the estimation times, $T_{MSM} / T_{ML}$,  are, respectively, \timeratiobcmfull, \timeratiobcmpartial, and \timeratiobcmnoisy, for the three scenarios respectively.
Aggregating the three scenarios, we have a median reduction of estimation time by $\timeratioaggregate \times$ (median estimation time of {\timepgabmaggregate}s for ML vs {\timemsmaggregate}s for MSM).

ML outperforms MSM in all the configurations of parameters tested.
In particular, ML always reduces the average error as the dataset size increases, both in terms of $T$, $m$, and $k$.
Conversely, MSM shows an unclear trend when varying these parameters: the error decreases as $T$ increases, is not affected by $k$, and has a bell shape when varying $m$.
Due to space constraints, we leave the exploration of this particular behavior of MSM to future research and show these results in the \suppmat.

\begin{figure*}
\centering
\includegraphics[width=.8\linewidth]{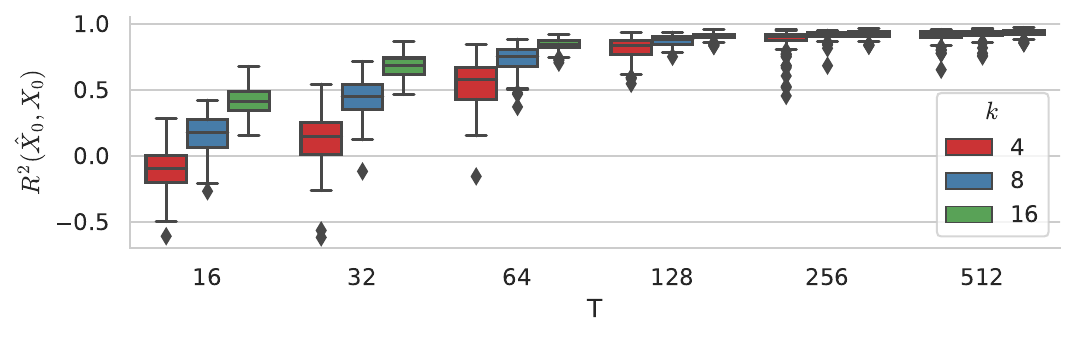}
\caption{
Distribution of $R^2$ scores of the estimate of the initial condition $\hat{X}_0$ with Maximum Likelihood as a function of the length of the data trace $T$ and the number of noisy observations of the opinions $k$.
The number of edges per timestep is $m=4$.
}
\label{fig:X0_boxplot}
\end{figure*}

One important advantage of ML is that it can also provide an estimate of the latent variable \xzero in \bcmnoisy.
\Cref{fig:X0_boxplot} displays the distribution of the coefficients of determination $R^2(\hat{\xzero}, \xzero)$ as a function of $T$ and $k$.
Besides the experiments with very small datasets, ML is able to estimate \xzero with a high level of accuracy.
Looking at experiments with $T \geq 64$, we obtain a median value of $R^2(\xzero)=\medianXRtwohighT$.
This result is particularly interesting since $\xzero \in [0,1]^{100}$ is high-dimensional, which indicates that this method could be employed for more complicated models with larger parameter spaces.
A comparison of the true values of \xzero and estimated ones $\hat{X}_0$ for an experiment with a median $R^2$ can be found in \suppmat.

\section{Discussion}

Our paper presents a comprehensive comparison between two distinct approaches to calibrating Agent-Based Models (ABMs) for opinion dynamics. 
We highlight several compelling factors that advocate for the adoption of a maximum likelihood (ML) approach over a simulation-based method (MSM) when calibrating an ABM.

\spara{Experimental evidence.}
Our experiments span a broad range of scenarios w.r.t. data availability.
The results clearly show the superiority of ML in terms of computational cost and estimation accuracy in all the contexts we considered.
The differences in the performances reach orders of magnitude.
The gap in performance persists as the dataset size increases. %

We compared the ML approach with a representative simulation-based method. 
We opted for MSM as it is the most popular and consolidated technique, and it captures most of the typical behaviors of its class of methods.
While more complex methods exist (see \Cref{sec:related}), there is no extensive assessment of their comparative performance, and thus it is hard to prefer one approach over another.
Additionally, several methods are restricted to a specific framework, and their adaptation to different models is non-trivial.

\spara{Theoretical robustness.}
The ML approach is grounded in a robust statistical framework that aligns with our specific problem. 
Rather than resorting to a trial-and-error methodology, we directly identify the parameters that most likely generated a given dataset. 
In contrast, the MSM method inherently relies on selected summary statistics and a chosen distance metric to characterize and compare simulated and observed time series. 
These selections wield significant influence over the calibration results, potentially discarding valuable information not encapsulated within these summaries.
Although there is potential to improve the choices of the summary statistics, distance measures, and sampling strategies, conducting an extensive optimization of the MSM for the presented scenarios is out of the scope of this paper.
Indeed, our point is that these choices will always be somewhat arbitrary, while using the data likelihood represents a principled approach to the issue.
Conversely, the ML approach casts the inference problem into an optimization task, which nevertheless needs careful hyperparameter tuning.
Fortunately, this process can leverage the extensive body of knowledge built by the machine learning community, which offers valuable insights and established methodologies.

\spara{Appropriatedness of the probabilistic framework.}
We used Probabilistic Graphical Models (PGMs) to formalize the rules of an Agent-Based Model (ABM) in a mathematical framework.
In itself, this formalization is already a noteworthy contribution, given the reproducibility issues that affect the ABM literature~\cite{wilensky2007making,fitzpatrick2019issues}.
Several factors support this choice:
\begin{squishlist}
    \item We have complete knowledge of the conditional probability distributions implied by the rules of the ABM.
    Therefore, it is natural to represent the causal connections via a graph of dependencies.
    \item The resultant PGMs often exhibit remarkable sparsity, mirroring the tendency of ABMs to involve only a handful of elements in causal relationships.
    This characteristic greatly simplifies the inference process for PGMs.
    \item PGMs can leverage the entire dataset to infer latent variables from the observed ones.
    This feature is missing in simulation-based approaches, which seed simulations with random initial conditions in an attempt to align with observed variables, essentially by chance.
\end{squishlist}

\spara{Estimated variables.}
In this work, we focused on the estimation of a specific parameter, $\epsilon$.
Notably, the limited dimensionality of the parameter space and the straightforward nature of the examined ABM should provide an optimal environment for the MSM approach. 
Despite these conditions, ML consistently outperforms MSM across all scenarios, even in the simplest, fully-observed one.
Additionally, the experiments in \bcmnoisy display that ML is able to effectively recover the initial conditions of the ABM in a 100-dimensional space.
These findings open the door to broader investigations involving higher-dimensional parameter spaces and more intricate ABMs.

\spara{Generalization and future work.}
A limitation of our work is the use of a single opinion dynamics model.
While our approach can be extended to more complex models, the process is not straightforward.
Indeed, one of the root causes of the difficulty in calibrating and validating ABMs is that their rules can quickly become very complex.
The generalizability of our results deserves further study..

However, there are reasons to be optimistic.
For instance, even though optimizing discrete probabilistic programs is still challenging, recent advances have been made in the area~\cite{arya2022automatic,lew2023adev}.
These capabilities would allow including a larger class of models with discrete parameters and variables, common in ABMs, within the maximum likelihood framework.
In addition, even when the likelihood function is not tractable exactly, approximate techniques such as continuous approximations or variational inference can be used to tackle complex models.
By connecting the world of ABM to probabilistic models and machine learning, we can hope to leverage the rapid advancements that these fields are experiencing.

\bibliographystyle{ACM-Reference-Format}
\bibliography{bibliography}

\end{document}

\maketitle

\section{Experiments}

\begin{figure*} [h]
\centering
\includegraphics[width=\linewidth]{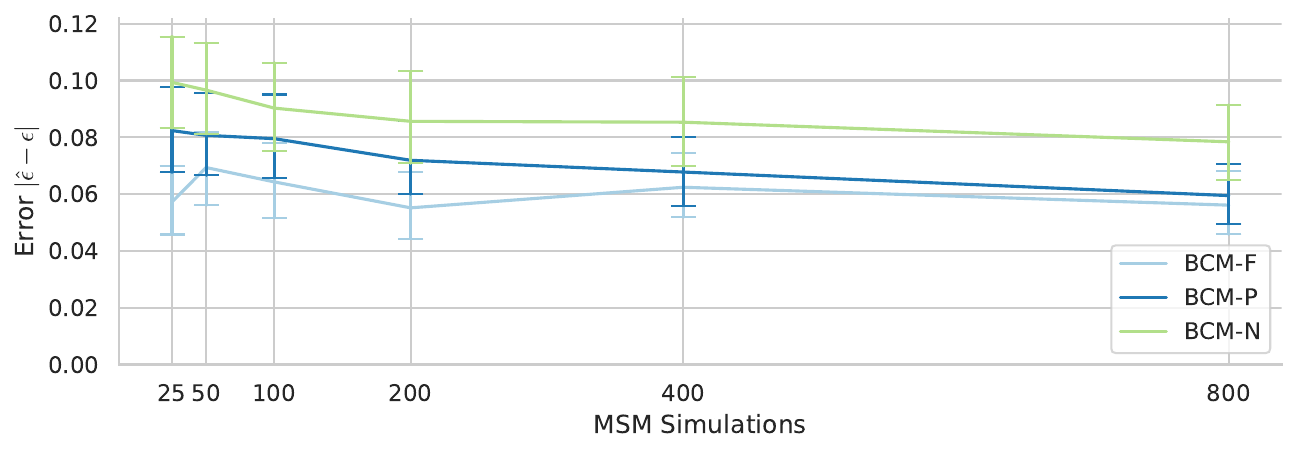}
\caption{
Comparison of the estimation error $\lvert \epsilon - \hat{\epsilon} \rvert$ with different number of simulations.
It is possible to notice a raise of the accuracy of the estimate as the number of simulations increases.
The plot aggregate the results of the three scenarios under study.
}
\label{fig:tuning_msm}
\end{figure*}

\spara{MSM parameter settings.}

\Cref{fig:tuning_msm} compares the error $\lvert \hat{\epsilon} - \epsilon \rvert$ of MSM as a function of the number of simulations.
For tuning the number of simulations, we repeat 90 simulations for each scenario, varying $T$ and the seed of the pseudorandom number generator.
As expected, we observe better performances as the number of simulations increases.
However, the estimation time grows linearly with the number of simulations.
Hence, we have to balance a time-accuracy trade-off.
From the plot we can observe a knee corresponding to \num{200} simulations, indicating that the estimation accuracy flattens after \num{200}.
For this reason, we run all the experiments with \num{200} simulations.

\clearpage

\spara{ML parameter settings.}

We use PyTorch, a Python framework that enables automatic differentiation, to maximize the objective functions resulting from each scenario.

We tune the hyperparameters by looking at all the possible combinations of values.
Notably, in \bcmnoisy, the optimization process involves both the estimation of target parameters $\epsilon$ and the latent variable \xzero.
The tuning involved the following hyperparameters:
\begin{squishlist}
    \item Optimizer: Adam, Nadam, Adagrad, RMSprop, SGD.
    \item Learning rate for $\epsilon$: 0.0001, 0.0005, 0.001, 0.005, 0.01, 0.05, 0.1.
    \item Learning rate multiplier for \xzero: 1, 10, 20, 50. This parameter is a multiplier for the learning rate of \xzero compared to the learning rate of $\epsilon$.
    \item Minibatch size: 16, 32, no-minibatching. This parameter controls the number of timesteps considered in the computation of the likelihood.
\end{squishlist}

For each scenario, we choose the optimizer that minimizes the average error $\lvert \hat{\epsilon} - \epsilon \rvert$.
The specific optimizers and their respective learning rates for each scenario are:
\begin{squishlist}
\item    \bcmfull: Nadam with a learning rate of 0.1.
\item    \bcmpartial: Adam with a learning rate of 0.05.
\item    \bcmnoisy: RMSprop with learning rates set to 0.05 for $\epsilon$ and set to 0.5 for \xzero (multiplier set to \num{10}).
\end{squishlist}
We do not use minibatching in any scenario.
All optimizers are stopped after \num{200} epochs if they have not converged yet.

\clearpage

\spara{Results.}

\begin{figure*} 
\centering
\includegraphics[width=1.\linewidth]{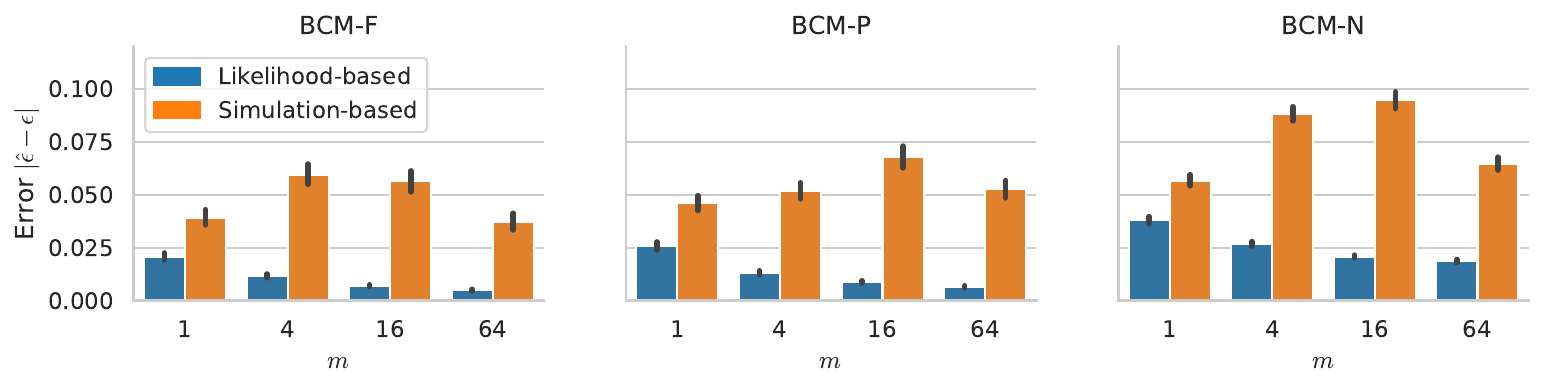}
\caption{
Average estimation error of $\epsilon$ by ML and MSM as a function of $m$ (number of edges per timestep.
Error bars represent the $95\%$ confidence intervals.
ML always produces significantly more accurate estimates than MSM.
ML estimates improve as $m$ increases.
This behavior is not observable in MSM, that consistently shows bell shapes, with lowest performances around middle values.
}
\label{fig:bar_edge_per_t}
\end{figure*}

\begin{figure*} 
\centering
\includegraphics[width=.35\linewidth]{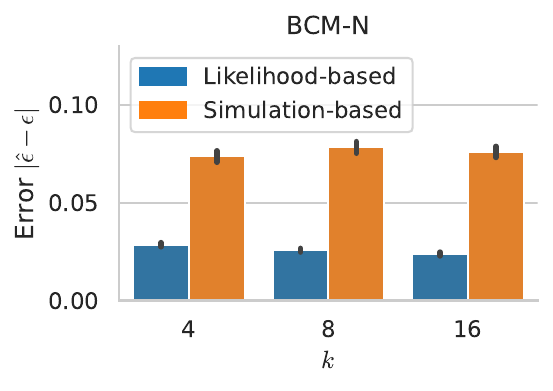}
\caption{
Average estimation error of $\epsilon$ by ML and MSM as a function of $k$ (number of noisy opinion proxies per timestep).
The plot refers only to \bcmnoisy, because $k$ is not present in the other scenarios.
The error bars represent the $95\%$ confidence intervals.
ML significantly improves the estimates w.r.t. MSM for every value of $k$.
The ML estimate improves as $k$ increases.
Instead, $\lvert \hat{\epsilon}_{MSM} - \epsilon \rvert$ does not show significant variations.
}
\label{fig:bar_evidences_per_t}
\end{figure*}

\Cref{fig:bar_edge_per_t} and \Cref{fig:bar_evidences_per_t} compare the estimation performances of ML and MSM as a function of $m$ and $k$.
In general, ML improves as the dataset size size increases.
This is true for $T$, $m$ and $k$.
Conversely, MSM shows diminished accuracy for the middle values of $m$ (\Cref{fig:bar_edge_per_t}).
Finally, MSM is not affected by the number of noisy opinions per timestep (\Cref{fig:bar_evidences_per_t}).
\Cref{table:bcmfull,table:bcmpartial,table:bcmnoisy} present the median estimation errors and median estimation time of the experiments for ML and MSM in detail.

\clearpage

\begin{figure*} 
\centering
\includegraphics[width=.4\linewidth]{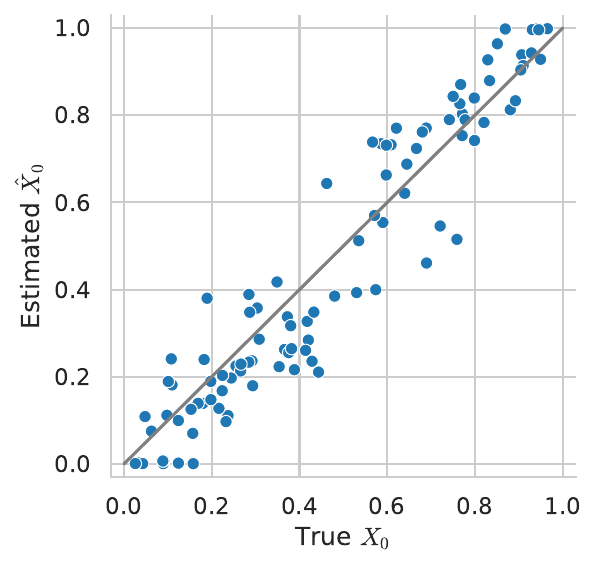}
\caption{
Comparison of the true $X_0$ with the estimated $\hat{X}_0$ of an experiment with a median value of $R^2$ (0.81).
}
\label{fig:realization_median_X0}
\end{figure*}

The optimization of the likelihood of \bcmnoisy involves the estimation both of $\hat{\epsilon}$ and $\hat{\xzero}$.
To assess the quality of our estimation, we compute the coefficient of determination $R^2(\hat{\xzero}, \xzero)$, defined as $R^2 = 1 - SS_{res} / SS_{tot}$, where $SS_{res}$ is the residual sum of squares, and $SS_{tot}$ is total sum of squares.
Intuitively, it measures the proportion of the variance of the data that is explained by the model.
In our experiments, we get a median $R^2(\hat{X}_0, X_0)$=\medianXRtwo.
This correspond to a median $MAE$=\medianXmae, where $MAE(\hat{X}_0, X_0) = \sfrac{1}{N} \sum_i \lvert X_0^i - \hat{X}_0^i\rvert$ (mean absolute error).

\Cref{fig:realization_median_X0} shows the comparison of true and estimated $X_0$ for a case with a median value of $R^2$.

\clearpage
\begin{table}[h]
\centering  
\caption{\bcmfull: full comparison of the median absolute errors ($\delta = \lvert \hat{\epsilon} - \epsilon \rvert$) and estimation time ($t$) between ML and MSM. Each value is the median over \numexperiments experiments.}

    \begin{tabular}{llrrrr}\toprule
$T$ & $m$ & $\delta_{ML}$ & $\delta_{MSM}$ &  $t_{ML}$ &  $t_{MSM}$ \\
\midrule
16  & 1  &        0.036 &         0.074 &   0.025 &    1.438 \\
    & 4  &        0.019 &         0.079 &   0.025 &    1.601 \\
    & 16 &        0.012 &         0.084 &   0.024 &    1.601 \\
    & 64 &        0.005 &         0.064 &   0.025 &    1.610 \\
\cmidrule(lr){1-6}
32  & 1  &        0.021 &         0.043 &   0.023 &    2.131 \\
    & 4  &        0.014 &         0.062 &   0.026 &    2.477 \\
    & 16 &        0.007 &         0.064 &   0.025 &    2.478 \\
    & 64 &        0.004 &         0.032 &   0.026 &    2.493 \\
\cmidrule(lr){1-6}
64  & 1  &        0.018 &         0.027 &   0.027 &    3.511 \\
    & 4  &        0.010 &         0.058 &   0.026 &    4.217 \\
    & 16 &        0.006 &         0.045 &   0.026 &    4.227 \\
    & 64 &        0.003 &         0.017 &   0.029 &    4.253 \\
\cmidrule(lr){1-6}
128 & 1  &        0.011 &         0.022 &   0.025 &    6.265 \\
    & 4  &        0.007 &         0.044 &   0.026 &    7.681 \\
    & 16 &        0.004 &         0.024 &   0.028 &    7.702 \\
    & 64 &        0.003 &         0.012 &   0.038 &    7.742 \\
\cmidrule(lr){1-6}
256 & 1  &        0.010 &         0.016 &   0.023 &   11.820 \\
    & 4  &        0.005 &         0.021 &   0.025 &   14.612 \\
    & 16 &        0.004 &         0.016 &   0.031 &   14.677 \\
    & 64 &        0.003 &         0.012 &   0.040 &   14.789 \\
\cmidrule(lr){1-6}
512 & 1  &        0.007 &         0.010 &   0.027 &   22.945 \\
    & 4  &        0.004 &         0.014 &   0.029 &   28.607 \\
    & 16 &        0.003 &         0.011 &   0.033 &   28.920 \\
    & 64 &        0.002 &         0.013 &   0.066 &   28.838 \\
\bottomrule
\end{tabular}

\label{table:bcmfull}

\end{table}

\clearpage

\begin{table}[h]
    \centering
\caption{\bcmpartial: full comparison of the median absolute errors ($\delta = \lvert \hat{\epsilon} - \epsilon \rvert$) and estimation time ($t$) between ML and MSM.
Each value is the median over \numexperiments experiments.
}
\begin{tabular}{llrrrr}
\toprule
$T$ & $m$  &   $\delta_{ML}$ & $\delta_{MSM}$ &  $t_{ML}$ &  $t_{MSM}$ \\
\midrule
16  & 1  &        0.035 &         0.076 &   0.185 &    2.088 \\
    & 4  &        0.020 &         0.080 &   0.177 &    2.093 \\
    & 16 &        0.011 &         0.096 &   0.172 &    2.089 \\
    & 64 &        0.007 &         0.090 &   0.186 &    2.101 \\
\cmidrule(lr){1-6}
32  & 1  &        0.029 &         0.041 &   0.297 &    3.443 \\
    & 4  &        0.015 &         0.053 &   0.302 &    3.465 \\
    & 16 &        0.010 &         0.073 &   0.320 &    3.455 \\
    & 64 &        0.006 &         0.051 &   0.305 &    3.478 \\
\cmidrule(lr){1-6}
64  & 1  &        0.021 &         0.030 &   0.560 &    6.146 \\
    & 4  &        0.011 &         0.042 &   0.534 &    6.191 \\
    & 16 &        0.007 &         0.063 &   0.532 &    6.171 \\
    & 64 &        0.003 &         0.039 &   0.555 &    6.216 \\
\cmidrule(lr){1-6}
128 & 1  &        0.019 &         0.031 &   1.086 &   11.614 \\
    & 4  &        0.007 &         0.034 &   1.036 &   11.699 \\
    & 16 &        0.006 &         0.044 &   1.043 &   11.615 \\
    & 64 &        0.004 &         0.027 &   1.031 &   11.727 \\
\cmidrule(lr){1-6}
256 & 1  &        0.010 &         0.021 &   1.963 &   22.300 \\
    & 4  &        0.006 &         0.021 &   1.988 &   22.457 \\
    & 16 &        0.004 &         0.023 &   1.979 &   22.417 \\
    & 64 &        0.002 &         0.028 &   1.930 &   22.620 \\
\cmidrule(lr){1-6}
512 & 1  &        0.009 &         0.018 &   3.939 &   43.786 \\
    & 4  &        0.005 &         0.017 &   3.801 &   44.190 \\
    & 16 &        0.003 &         0.021 &   3.862 &   44.113 \\
    & 64 &        0.002 &         0.019 &   3.953 &   44.508 \\
\bottomrule
\end{tabular}
\label{table:bcmpartial}

\end{table}

\clearpage
\begin{table}[h]
\centering

\caption{\bcmnoisy: full comparison of the median absolute errors ($\delta = \lvert \hat{\epsilon} - \epsilon \rvert$) and estimation time ($t$) between ML and MSM.
Each value is the median over \numexperiments experiments.
}

\begin{minipage}{0.5\textwidth}
\begin{tabular}{lll rrrr}
\toprule
$T$ & $m$ & $k$ &  $\delta_{ML}$ & $\delta_{MSM}$ &  $t_{ML}$ &  $t_{MSM}$ \\
\midrule
16  & 1  & 4  &        0.060 &         0.074 &   1.055 &    2.768 \\
    &    & 8  &        0.059 &         0.079 &   1.111 &    2.758 \\
    &    & 16 &        0.057 &         0.075 &   0.870 &    2.750 \\
\cmidrule(lr){2-7}
    & 4  & 4  &        0.045 &         0.106 &   0.838 &    2.899 \\
    &    & 8  &        0.035 &         0.131 &   0.903 &    2.900 \\
    &    & 16 &        0.037 &         0.115 &   0.796 &    2.901 \\
\cmidrule(lr){2-7}
    & 16 & 4  &        0.026 &         0.150 &   0.754 &    2.902 \\
    &    & 8  &        0.026 &         0.136 &   0.734 &    2.903 \\
    &    & 16 &        0.022 &         0.140 &   0.730 &    2.903 \\
\cmidrule(lr){2-7}
    & 64 & 4  &        0.014 &         0.103 &   0.724 &    2.910 \\
    &    & 8  &        0.017 &         0.128 &   0.662 &    2.926 \\
    &    & 16 &        0.017 &         0.144 &   0.657 &    2.926 \\
\cmidrule(lr){1-7}
32  & 1  & 4  &        0.045 &         0.056 &   1.612 &    3.663 \\
    &    & 8  &        0.032 &         0.060 &   1.402 &    3.653 \\
    &    & 16 &        0.035 &         0.051 &   1.192 &    3.648 \\
\cmidrule(lr){2-7}
    & 4  & 4  &        0.036 &         0.078 &   1.495 &    3.954 \\
    &    & 8  &        0.022 &         0.107 &   1.406 &    3.955 \\
    &    & 16 &        0.039 &         0.083 &   1.216 &    3.954 \\
\cmidrule(lr){2-7}
    & 16 & 4  &        0.018 &         0.126 &   1.286 &    3.961 \\
    &    & 8  &        0.017 &         0.144 &   1.238 &    3.960 \\
    &    & 16 &        0.022 &         0.142 &   1.247 &    3.960 \\
\cmidrule(lr){2-7}
    & 64 & 4  &        0.018 &         0.054 &   1.458 &    3.969 \\
    &    & 8  &        0.018 &         0.072 &   1.374 &    3.988 \\
    &    & 16 &        0.018 &         0.064 &   1.260 &    3.989 \\
\cmidrule(lr){1-7}
64  & 1  & 4  &        0.041 &         0.035 &   2.231 &    5.455 \\
    &    & 8  &        0.045 &         0.046 &   2.145 &    5.446 \\
    &    & 16 &        0.028 &         0.039 &   1.896 &    5.443 \\
\cmidrule(lr){2-7}
    & 4  & 4  &        0.022 &         0.076 &   2.553 &    6.059 \\
    &    & 8  &        0.031 &         0.095 &   2.150 &    6.061 \\
    &    & 16 &        0.022 &         0.084 &   2.479 &    6.060 \\
\cmidrule(lr){2-7}
    & 16 & 4  &        0.020 &         0.079 &   2.503 &    6.069 \\
    &    & 8  &        0.023 &         0.094 &   2.517 &    6.071 \\
    &    & 16 &        0.015 &         0.080 &   2.831 &    6.070 \\
\cmidrule(lr){2-7}
    & 64 & 4  &        0.020 &         0.035 &   3.114 &    6.089 \\
    &    & 8  &        0.017 &         0.044 &   2.811 &    6.111 \\
    &    & 16 &        0.016 &         0.036 &   2.786 &    6.112 \\
\bottomrule
\end{tabular}

\end{minipage}%
\begin{minipage}{0.5\textwidth}
\begin{tabular}{lllrrrr}
\toprule
    $T$ & $m$ & $k$ &   $\delta_{ML}$ & $\delta_{MSM}$ &  $t_{ML}$ &  $t_{MSM}$ \\
\midrule
128 & 1  & 4  &        0.037 &         0.034 &   4.172 &    9.049 \\
    &    & 8  &        0.024 &         0.034 &   3.991 &    9.039 \\
    &    & 16 &        0.018 &         0.036 &   3.915 &    9.033 \\
\cmidrule(lr){2-7}
    & 4  & 4  &        0.021 &         0.068 &   5.045 &   10.277 \\
    &    & 8  &        0.017 &         0.080 &   4.745 &   10.288 \\
    &    & 16 &        0.014 &         0.075 &   4.491 &   10.282 \\
\cmidrule(lr){2-7}
    & 16 & 4  &        0.017 &         0.051 &   5.642 &   10.304 \\
    &    & 8  &        0.016 &         0.041 &   5.166 &   10.300 \\
    &    & 16 &        0.014 &         0.054 &   5.893 &   10.297 \\
\cmidrule(lr){2-7}
    & 64 & 4  &        0.018 &         0.031 &   6.051 &   10.348 \\
    &    & 8  &        0.016 &         0.035 &   5.937 &   10.371 \\
    &    & 16 &        0.016 &         0.029 &   5.487 &   10.364 \\
\cmidrule(lr){1-7}
256 & 1  & 4  &        0.023 &         0.026 &   8.533 &   16.249 \\
    &    & 8  &        0.017 &         0.033 &   8.376 &   16.231 \\
    &    & 16 &        0.013 &         0.031 &   7.743 &   16.238 \\
\cmidrule(lr){2-7}
    & 4  & 4  &        0.015 &         0.038 &  10.554 &   18.720 \\
    &    & 8  &        0.014 &         0.037 &  10.256 &   18.744 \\
    &    & 16 &        0.011 &         0.039 &   9.511 &   18.735 \\
\cmidrule(lr){2-7}
    & 16 & 4  &        0.018 &         0.033 &  11.705 &   18.765 \\
    &    & 8  &        0.015 &         0.032 &  12.910 &   18.771 \\
    &    & 16 &        0.013 &         0.040 &  13.244 &   18.771 \\
\cmidrule(lr){2-7}
    & 64 & 4  &        0.013 &         0.031 &  11.316 &   18.836 \\
    &    & 8  &        0.012 &         0.022 &  11.141 &   18.878 \\
    &    & 16 &        0.010 &         0.027 &  11.665 &   18.883 \\
\cmidrule(lr){1-7}
512 & 1  & 4  &        0.015 &         0.022 &  24.848 &   30.796 \\
    &    & 8  &        0.012 &         0.023 &  22.043 &   30.756 \\
    &    & 16 &        0.009 &         0.019 &  23.562 &   30.773 \\
\cmidrule(lr){2-7}
    & 4  & 4  &        0.014 &         0.024 &  28.525 &   35.784 \\
    &    & 8  &        0.013 &         0.026 &  27.683 &   35.736 \\
    &    & 16 &        0.009 &         0.031 &  24.248 &   35.754 \\
\cmidrule(lr){2-7}
    & 16 & 4  &        0.015 &         0.022 &  30.236 &   35.888 \\
    &    & 8  &        0.012 &         0.026 &  31.554 &   35.792 \\
    &    & 16 &        0.010 &         0.022 &  32.369 &   35.808 \\
\cmidrule(lr){2-7}
    & 64 & 4  &        0.009 &         0.023 &  25.224 &   36.063 \\
    &    & 8  &        0.009 &         0.019 &  23.575 &   36.012 \\
    &    & 16 &        0.007 &         0.017 &  21.827 &   36.080 \\
\bottomrule
\end{tabular}

\end{minipage}

\label{table:bcmnoisy}

\end{table}